\documentclass[preprint2]{aastex}
\usepackage{amsmath, flmanyeq}

\newcommand{\be}{\begin{equation}}
\newcommand{\ee}{\end{equation}}

\newcommand{\boldx}{\mathbf x}
\newcommand{\boldk}{\mathbf k}

{

}

\shorttitle{Quaquaversal tiles: analytical pre-initial conditions}
\shortauthors{Various authors}
\slugcomment{To be submitted to ApJL}

\begin{document}

\title{An alternative to grids and glasses:\\
Quaquaversal pre-initial conditions for N-body simulations}

\author{Steen H. Hansen$^{1,2}$, Oscar Agertz$^1$, Michael Joyce$^3$,\\
Joachim Stadel$^1$, Ben Moore$^1$, Doug Potter$^1$}

\affil{$^1$ University of Zurich, Winterthurerstrasse 190,
8057 Zurich, Switzerland}

\affil{$^2$ Dark Cosmology Center, Niels Bohr Institute, University
of Copenhagen, \\Juliane Maries Vej 30, 2100 Copenhagen, Denmark}

\affil{$^3$ Laboratoire de Physique Nucl\'eaire et des Hautes
Energies,\\ Universit\'e Pierre et Marie Curie-Paris 6, UMR 7585,\\
Paris, F-75005 France}




\begin{abstract}
N-body simulations sample their initial conditions on an initial
particle distribution, which for cosmological simulations is usually a
glass or grid, whilst a Poisson distribution is used for galaxy
models, spherical collapse etc. These pre-initial conditions have
inherent correlations, noise due to discreteness and preferential
alignments, whilst the glass distribution is poorly defined and
computationally expensive to construct.  We present a novel particle
distribution which can be useful as a pre-initial condition for N-body
simulations, using a simple construction based on a ``quaquaversal''
tiling of space. This distribution has little preferred orientation
(i.e. is statistically isotropic), has a rapidly vanishing large scale
power-spectrum ($P(k) \sim k^{4}$), and is trivial to create. 
It
should be particularly useful for warm dark matter and cold collapse
simulations.
\end{abstract}




\section{INTRODUCTION}

Numerical simulations have become a very powerful tool for investigating 
non-linear gravitational phenomenon, such as
understanding the evolution and properties of cosmological
structures. Since the simulations contain a rapidly increasing number
of particles, and probe structures on ever smaller scales, it is
timely to address some of the fundamental aspects of the initial
conditions used in these simulations.

One such aspect is that the cosmological standard model assumes that
the early Universe is statistically isotropic. This is in contrast
with the usual choice of {\em pre-initial} conditions for N-body
simulations, most often given by placing particles on a uniform grid,
which is intrinsically anisotropic.  While it has been shown that this
anisotropy can produce non-physical effects in simulations, such
effects are difficult to quantify. For cold dark matter simulations it
is thought that the physically relevant correlations quickly grow and
dominate over fluctuations due to discreteness. The same is not true
for warm dark matter simulations for example.

To address this and related questions numerically it is useful to 
have alternative pre-initial conditions (PreICs). We construct a novel
pre-initial condition by making use of a tiling of three dimensional
space called the {\em quaquaversal tiling} \citep{conrad}.
We will show that this new particle distribution,
which is statistically isotropic 
(i.e. has little intrinsic directionality), has mass fluctuations which 
decay as rapidly as in a grid. At
the same time it is a deterministic structure, and can be 
trivially generated. 
We present a C-code for the generation of these
structures, which can be downloaded from {\tt
http://krone.physik.unizh.ch/{}\~{}hansen/qua/}.

\section{EXISTING PRE-INITIAL CONDITIONS}

Let us briefly recall the steps in any cosmological simulation. 
1) First one chooses PreICs, which most often is a regular grid, i.e. the
particles are placed on a lattice.  2) One then imprints
a power-spectrum onto these particles, by applying an 
appropriate displacement field specified through the Zeldovich
approximation.  3) Then one runs the cosmological code, taking care
of the many numerical issues related to convergence (softening,
time stepping etc).  The present paper focuses solely on the first
step, namely the setting up of the {\em pre-initial} condition.

There are three PreICs 
which are regularly used in the literature.  
The first,
which is the standard choice, places the particles on a regular
grid. This is a very well tested method, which most likely produces
the correct growth of long range large scale fluctuations 
\citep{efst85}.
However, it is unknown how
much the grid affects small scale structures.  One explicit
example in which such effects have been observed to be important
are in simulations with warm dark matter (WDM). In the WDM case 
the thermal motion induces a free streaming, which erases
structures on small scales. In a paper by \cite{bode} it was suggested
that WDM might have the novel property of creating structures along a
cosmic web, below the cut-off frequency of the power spectrum. 
\cite{bode} took great care in testing for a large range
of known numerical issues, and concluded that the effect observed
was real and physical.  
It was later shown that these small scale structures were spurious.
In the paper by \cite{gsl} two almost identical simulations were 
performed, differing only in the {\em pre-initial} conditions: one was 
a grid, the other was a glass (we will discuss {\em glass} initial 
conditions below). It
was shown from the results of these simulations \citep{gsl}, that the
conclusions reached in \cite{bode} were incorrect precisely because
of effects coming from the pre-initial particle distribution: the bead-like structures 
along the filaments observed were virtually absent in  the simulation 
with a  pre-initial glass. We emphasise that this is just an illustration 
of how difficult WDM simulations can be. Similar difficulties have been
discussed in the context of cold dark matter (CDM) simulations
e.g. starting from specific configurations \citep{melott},
and at early times \citep{joyceetal-DISC}, it remains unclear how 
important such effects are in real simulations.

The second PreICs which are often used is the {\em
glass}. The idea is to evolve a set of particles, initially
randomly distributed in a box, under negative gravity, i.e. Coulomb
forces, until one reaches a 
configuration in which the force on each 
particle is extremely small \citep{white}. While this appears simple
at first sight, there are a range of well known practical problems.
Firstly, starting from the random configuration, the particles
stream towards a lower potential, but gain kinetic energy
which makes them oscillate about the minimum of their local
potential. To reduce the associated Poisson noise, one needs
to damp these velocities. If one does so by reducing the particle 
velocities at a given time, then a large fraction of the particles 
will lie far from the minimum of the potential and little 
is gained.  If one applies a more continuous damping of the particle 
velocities, then either the small scale fluctuations are
erased (if the damping is large), or the glass takes an unreasonably
long time to create (if the damping is small). 
If one uses the method  of simulated annealing, repeatedly heating up and 
cooling down the system, the creation of the glass becomes 
computationally very expensive. A major difficulty is that
the final configuration is not unique, and indeed that one does not 
have a well defined criterion for determining when an optimal 
configuration has been reached. We note in
this respect (see \cite{gabriellietal-OCP}) that the system which is 
simulated  is just a damped variant of the ``one component plasma'' 
(i.e. point particles interacting through Coulomb forces), which
is known \citep{carr1961} to undergo a transition to a body centered cubic
lattice configuration at low temperature.  The glass is
presumably a transient to such a grid type configuration,
but it is not known what the relevant time scales are.

A third possibility, and one that is frequently used to create 
equilibrium and non-equilibrium halo models is to simply select 
random positions for particles. In this case there is 
Poisson noise at all scales. If one tries to simulate a ``cold collapse''
using such initial conditions then the growth of small scale structures can 
clearly be seen which will affect the virialisation of the final structure.

In summary the {\em grid} PreICs are trivial to create and has
vanishing power-spectrum (below the Nyquist frequency). It has,
however, a strong orientation and power on the scale of the grid
spacing. The {\em glass} PreICs have, in principle, little preferred
orientation, and has a rapidly decreasing power spectrum of density
fluctuations, with $P(k) \sim k^4$ \citep{smithetal}, at large
scales. These latter configurations are, however, not clearly defined
and they are computationally expensive to create. Note, however,
that the glass PreICs still suffer from large scale anisotropy as long
as periodic boundary conditions are used. Such large scale anisotropies
are evidently also present in the new PreIC to be 
discussed below. The {\em random} PreIC 
has no orientation, but it has significant 
intrinsic power on all
scales ($P(k)={\rm constant})$).  We will now present a novel PreIC
which is clearly defined and easy to generate, has a large scale power
spectrum vanishing as in the glass configurations ($P(k) \sim k^4$),
and has no preferred orientation.

\begin{figure}[htb]
	\centering
	\includegraphics[angle=0,width=0.4\textwidth]{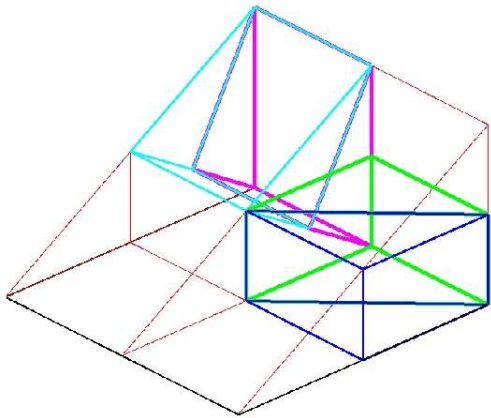}
	\caption{The quaquaversal tiling after one step. The thick
lines (coloured) show the rotated smaller triangles, which are rotated by
$2\pi/3$ and $\pi/2$ respectively.}
	\label{fig:qua}
\end{figure}

\section{CONSTRUCTING A QUAQUAVERSAL TILING}

The quaquaversal tiling \citep{conrad} is a hierarchical tiling of 
3 dimensional space, based on a triangular prism that is repeatedly 
rotated about orthogonal axes by angles $2\pi/3$ and $\pi/2$. 

The principle of our construction of the PreICs is
simple. The quaquaversal tiling defines a division of space into {\it
equal volume} cells. The particle distribution obtained
by assigning a particle of the same mass to each cell is highly 
uniform: the only source of fluctuations is the redistribution 
of the mass at the scale of the cell. Indeed a well known argument,
due to Zeldovich \citep{zeldovich, zeldovich-novikov}, shows that 
such a local mass conservation constraint should lead to a power 
spectrum with the behaviour $P(k) \sim k^2$ at small $k$. If one
adds the further constraint that the {\em centre of mass} is locally
conserved one expects to obtain $P(k) \sim k^4$ at small $k$. Here we 
will impose this constraint by placing the particle in each tile at
its centre of mass. Furthermore, the tiling has little preferred 
orientation, a desirable property which will be inherited by the 
particle distribution.

Consider a triangular tile, made from a 1, $\sqrt{3}$, 2 right-angle
triangle, with depth 1/2. This tile can be decomposed into 4 identical
tiles, all with exactly the same properties as the original ``parent'' 
tile, by
placing 3 lines from the center of the long side (length 2), to the
centers of the other 2 sides (length 1 and $\sqrt{3}$) and to the
right angle. Then all lines are extended in depth (in the 3rd
dimension).  
One can now choose one of two possibilities, either to rotate the two
triangles at the short axis by $2\pi/3$ about an axis in the depth
dimension, or to rotate the two triangles touching the right angle by
$\pi/2$ about an axis in the y-dimension (see
figure~\ref{fig:qua}). Finally, one places two tiles next to each
other, each with their choice of rotation.  For further detail, see
\cite{conrad}.

After this first tiling, we are left with 8 triangles, each identical
to the original parent triangle, but a factor 8 smaller in
volume.  (The angles of the individual triangles are the same, only the
sides are smaller by a factor 2).  We can now repeat this process again
for each triangle, giving us 64 identical triangles. After $N$
such iterations we will have $8^N$ identical triangles, with the same
properties as the original parent triangle.  

To obtain our configuration, we then place a particle in the center 
of volume of each triangle. Finally we place two parent tiles on top of 
each other, to form a rectangular box with sides 1,1,$\sqrt{3}$. 
This final distribution, which is our new PreIC to be used in
simulations, thus contains $2\times 8^N$ particles. From now on, we 
will refer to this kind of particle distribution as a \emph{Q-set}. For
simulations where periodic boundary conditions are needed, one can
only make Q-sets with $16,128, \cdots, 4.2M, 33.6M, \cdots$
particles. It is proved in \cite{conrad} that, in the limit 
of an infinite number of iterations, the orientations of
the tiles are essentially random (uniform in SO(3)). 
We infer that our distribution (with finite, but 
large, $N$) will have little directionality 
(see discussion in section \ref{sec:isotropy}).

Naturally one can imagine similar constructions
based on other tilings. In this paper, however, 
we will focus on this simple quaquaversal structure.

\subsection{COMPUTER CODE}

We will briefly describe the idea behind this code.  Each triangle is
uniquely defined through the definition of the spatial position (in
3-d space) of 4 corners (3 corners would
suffice, of course). 
Thus, the original parent tile is defined through 4 different
$3-$vectors.  This is expressed through one 12-vector, $\vec V$.  The
8 sub-tiles in the next level of tiling, $n+1$, are defined through a
constant matrix, ${\cal M}$, applied to each vector at level $n$ such
that
\be \vec V_i(n+1) = {\cal M} \vec V(n) \, ,
\ee
where $i=1, \cdots, 8$ define the 8 new sub-tiles. Thus the constant
matrix ${\cal M}$ has 128 entries, and it uniquely defines each tiling
step.

The code recursively applies ${\cal M}$ to the vectors (it recursively
tiles each subtile), until the desired level, $N$, has been
reached. Then a particle is placed in the center of volume of that
triangle, and that point is written to a file. Constructed in this way
the code is very fast and requires very little memory.

In our implementation the matrix ${\cal M}$ is constant. This implies that the
specific rotations by fractions of $\pi$ are identical at each level
of refinement. It would be possible to generalize the procedure, by
allowing the rotations to differ at each level, e.g. sometimes rotate
by $4\pi/3$ instead of $2\pi/3$. In that way one could achieve a
slightly higher level of isotropization.

We will now consider the properties of the particle distribution in this
rectangular box, in terms of mass variance and power spectrum.

\section{STATISTICAL PROPERTIES}

\subsection{Mass Variance $\sigma_M^2(R)$}

Let us first analyse the amplitude of mass
fluctuations in a sphere of radius $R$ with respect to the average
mass. If $M(R)$ is the mass (for a discrete distribution, the number of
particles) inside a sphere of radius $R$, the
normalised mass variance is defined as
\begin{equation}
\sigma_M^2(R)=      
\frac{\langle M(R)^2 \rangle - \langle M(R) \rangle^2}{\langle M(R) \rangle^2},
\end{equation}
where the brackets indicate an ensemble average. For a distribution
like ours (or e.g. a grid) one can define the ensemble average as the
average over random positions of the initial box. Such an
ensemble average definition is equivalent to a spatial average, with
the mass variance then given as the infinite volume limit of the 
estimator defined below.

\begin{figure}[t]
	\centering
	\includegraphics[angle=0,width=0.48\textwidth]{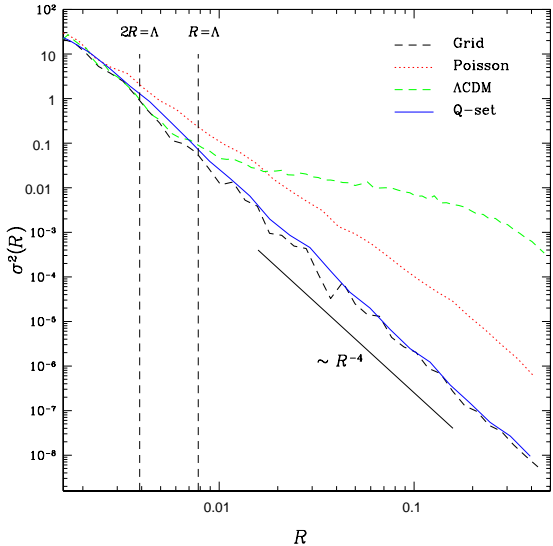}
	\caption{The mass variance as function of radius. The correct
(analytical) behaviour of the test structures are found. The Poisson
(red, dotted) has $\sigma^2 \sim R^{-3}$, the grid (black,
long-dashed) has $\sigma^2 \sim R^{-4}$ turning into $R^{-3}$ for distances
much below the interparticle distance. The $\Lambda$CDM (green,
dashed) has the correct behaviour until scales where the interparticle
distance approaches the grid size. The Q-set (blue, solid line) has
the same behaviour as a grid, namely $\sigma^2 \sim R^{-4}$ on all
scales down to the Poissonian turn over.}
	\label{fig:sigma}
\end{figure}

For our mass variance calculations we have used the 
simple estimator 
\begin{equation}\label{sigmaestimator}
 \sigma_{M, {\rm est}}^2(R) = \frac{1}{\langle N_r\rangle^2}
                              \sum_{i=1}^{N_s} \frac{(N_i(R) \ -
                              \langle N_r\rangle)^2}{N_s - 1} \, ,
\end{equation}
where $N_i(R)$ is the number of particles in the $i$th of $N_s$
randomly thrown spheres, constrained to be inside the sample
volume. $\langle N_r \rangle$ is the mean number of particles
in such a sphere, given exactly by 
$\langle N_r\rangle= \frac{4\pi r^3}{3 v}$ where
$v$ is the volume per particle (i.e. the volume of
a single tile in the Q-set).

We apply this estimator to a grid, a Poisson distribution, a typical
$\Lambda$CDM initial condition ($z=70$), and a Q-set. We have used
$128^3$ particles in a cube of side unity for the first three, while
the Q-set employed seven tiling levels, which gives $2\times8^7= 2
\times 128^3$ particles in a rectangular box. The dimension of
the Q-set box is chosen so that the mean density is equal to that of
the other distributions. This is a convenient choice for comparison
of the results, as it is the mean particle density $n_0$ which fixes
the asymptotic level of the Poisson variance at small scales in any
point distribution, with $\sigma_M^2(R \rightarrow 0) = \frac{1}{n_0
V_{\rm{s}}(R)}$ where $V_{\rm{s}}=4\pi R^3/3$ (see e.g. \cite{joyce}).
Several tests confirm that our estimator calculates the correct
spatial properties (see further discussion below). We see (figure
\ref{fig:sigma}) that the mass variance of a Q-set has the same
behaviour as that of a grid with $\sigma^2 \sim R^{-4}$ above the
interparticle distance.  The mean interparticle distance,
$\Lambda=1/128$, for the structures with $128^3$ particles is also
shown in the figure. Note that this is exactly the length of the
shortest side of a tile in the level seven quaquaversal tiling used
for the Q-set considered here.

\subsection{Power Spectrum $P(k)$}
The power spectrum is the primary statistical tool used to characterise
fluctuations in cosmology. It is defined as
\begin{equation}
\label{powerspectrum}
\label{powerpsec}
P({\mathbf {k}}) = \lim_{V \rightarrow \infty} \frac{1}{V} \langle|\hat\delta(\boldk)|^2 \rangle \, ,
\end{equation}
where $\hat\delta(\boldk)$ is the Fourier transform of the  density 
fluctuation field $\delta(\boldx)=(\rho(\boldx)-\rho_0)/\rho_0$. 
In the case of a discrete distribution (i.e. of point particles) these 
quantities simply become 
\begin{manyeqns}
\label{deltax}
\delta(\boldx) &=& \frac{V}{N_{\rm{p}}} \sum \delta_{\rm{D}}(\boldx-\boldx_{\rm
p}) -1 \, , \\ \hat \delta(\boldk) &=& \frac{V}{N_{\rm{p}}} \sum e^{-i \boldk \cdot
\boldx_{\rm p}} \qquad (\boldk \ne 0) \, , \label{deltak}
	\end{manyeqns}
where $\boldx_{\rm p}$ is the location of each particle and
$\delta_{\rm{D}}$ is the Dirac delta function. 

To estimate the power spectrum we have used the ``brute force"
method i.e. we calculate it directly from the formula one obtains 
by substituting Eq.~(\ref{deltak}) in Eq.~(\ref{powerspectrum}), 
{\it without} the infinite volume limit and ensemble average.
Our finite volume $V$ is thus a rectangular box with sides $L_i$,
and we assume periodic boundary conditions so that $\boldk$ in 
the Fourier  sums take the values
$\boldk = 2\pi \left( n_x/L_x, n_y/L_y, n_z/L_z \right) $ where
$n_i$ are integers. We obtain $P(k)=P(|\boldk|)$ by averaging
over a bin of finite width around $k=|\boldk|$~\citep{sirko}.
It is important to take care in the interpretation
of the large scale (i.e.  small $k$) modes, which will
be affected both by under-sampling and contaminated by
the boundary conditions (which systematically suppress
power at small $k$).  
With this simple estimator, however, we do not have to worry about 
the effect of assignment function, which typically causes problems 
in FFT methods \citep{jing}.

As for the mass variance we calculate the power spectrum for a grid, a Poisson distribution, a typical $\Lambda$CDM initial 
condition, and a Q-set. This time we use a smaller number of particles 
in order to facilitate the more computationally demanding procedure 
of calculating $P(k)$: they all have $32^3$ particles except for the 
Q-set which has $N=5$, and therefore again has twice as many 
points ($32^3=8^5$). Likewise for the mass variance, this makes the
the Poisson level ($V/N_{\rm{p}}$) the same for all 
the distributions considered.

\begin{figure}[htb]
	\centering
	\includegraphics[width=0.48\textwidth]{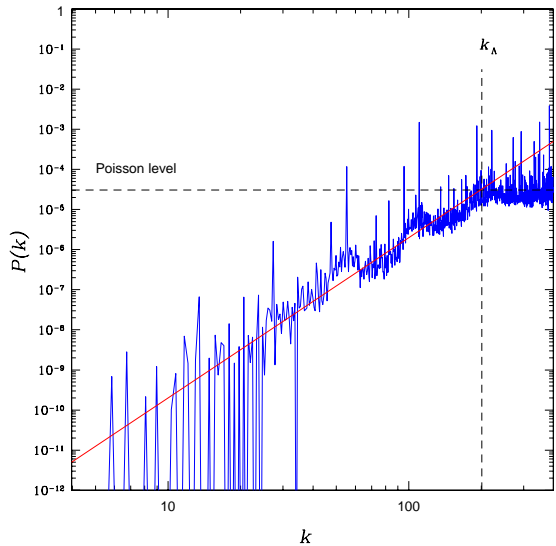}
	\caption{Detailed plot of the power spectrum for a the Q-set. We have used a very fine binning to capture the $k^4$ slope as well as the  characteristic peaks of this PreIC.}
	\label{fig:qpower}
\end{figure}

We see in figure \ref{fig:qpower} that the Q-set has the anticipated 
$P(k)\sim k^4$ behaviour up to the wave vector for the mean interparticle separation, $k_{\Lambda}=2\pi/\Lambda$, where it flattens on average, to the Poisson level. We have used a very fine binning in order to capture this characteristic slope as well as the peaks arising from typical interparticle distances. 
We note that peaks appear at frequencies 
of $2^{-n}k$, where $n$ has integer values. It
can be shown (see ~\cite{radin}) that in hierarchical structures, a
feature in real space between two radii, $r_1$ and $r_2$, also must
appear between the radii $\kappa r_1$ and $\kappa r_2$, where
$\kappa=2$ for a quaquaversal structure. 

Consistent with this interpretation we note that adding random 
perturbations to the particles, hence creating a shuffled 
Q-set, gives $P(k)\sim k^2$ with peaks of diminished amplitude.

\begin{figure}[htb]
	\centering
	\includegraphics[width=0.48\textwidth]{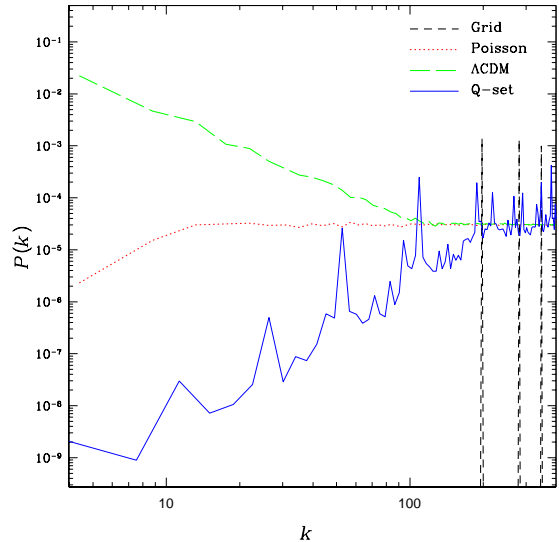}
	\caption{The power spectra for a Q-set (solid, blue), 
a grid (short-dashed, black), a Poisson distribution (dotted, 
red) and a $\Lambda$CDM initial condition (long-dashed, green) . The 
Q-set shows a $k^4$ behaviour up to the Poisson level. The power 
spectrums has characteristic peaks appearing, as discussed in the text, at $k/2,k/4,k/8\ldots\,$ 
due to the hierarchical properties of the Q-set.}
	\label{fig:power}
\end{figure}

A comparison between the different distributions can be seen in figure
\ref{fig:power}. Note that the grid has zero power up $k_\Lambda$
where it is strongly peaked and that all distributions, on average,
reach the Poisson level at this wave number.


\subsection{Relations between  $\sigma_M^2(R)$ and $P(k)$}
$\sigma^2(R)$ and $P(k)$ are related by  
the standard expression
\begin{equation}\label{sigmarPK}
\sigma_M^2(R) = \frac{1}{2\pi^2} 
\int_0^{+\infty} P(k) \hat{W}^2(kr) k^2 dk \, ,
\end{equation}
where $\hat{W}^2(kr)$ is the Fourier transform of the spherical top
hat window function.

By studying Eq.~(\ref{sigmarPK}) for a power spectrum of the
form $P(k \rightarrow  0)\sim k^n$ one finds ~\citep{joyce}
that for $R \rightarrow \infty$
\begin{equation} 
\label{pk-sr}  
\sigma_M^2(R) \sim   
\left\{ \begin{array}{lll} 
 1/R^{3+n} \; \;  \mbox{if} \;\; n<1\\   
\log(R)/R^{4} \;\; \mbox{if} \;\;  {n=1}  \\  
1/R^{4}  \;\; \mbox{if} \;\;  {n>1} 
\end{array}  
\right. 
\end{equation}

Two particular and simple examples which are useful reference
points against which to gauge new distributions are:
\begin{itemize}
\item Poisson: $\sigma_M^2(R)\sim R^{-3}$, $P(k) \sim V/N_{\rm p}
\sim const$
\item Shuffled Lattice: $\sigma_M^2(R) \sim R^{-4}$, $P(k) \sim k^2$
\end{itemize}

The numerical results for the mass variance and power spectrum
in the previous two sections are all in line with these 
analytic results, notably our new Q-set has 
$\sigma_M^2(R)\sim R^{-4}$ and $P(k) \sim k^4$. This (large scale)
behaviour makes the Q-set a member of a group of systems
which have been termed \emph{super-homogenous} 
~\citep{joyce, gabriellietal-OCP} or \emph{hyperuniform} 
~\citep{torquato}.
Such distributions, defined by the property
$P(k \rightarrow 0)=0$, are characterised in real space
(cf. Eq.~(\ref{pk-sr}) above)  by the asymptotic
behaviour of their variance, $\sigma^2(R) \sim 1/R^{m}$ with
$d<m<d+1$ where $d$ is the spatial dimension. This quantity
thus decays \emph{faster} than in a Poisson point process
($m=d$),  and in the cases we have considered attains the 
behaviour ($m=d+1$). This is in fact the fastest possible 
decay of this quantity for either point \citep{beck} or
continuous \citep{joyce} mass distributions.
We note that the glass PreIC also belong to
this class, as it shares this same behaviour of the 
variance and power spectrum as the Q-set we have 
introduced and analysed.

\begin{figure}[htb]
	\centering
	\includegraphics[width=0.48\textwidth]{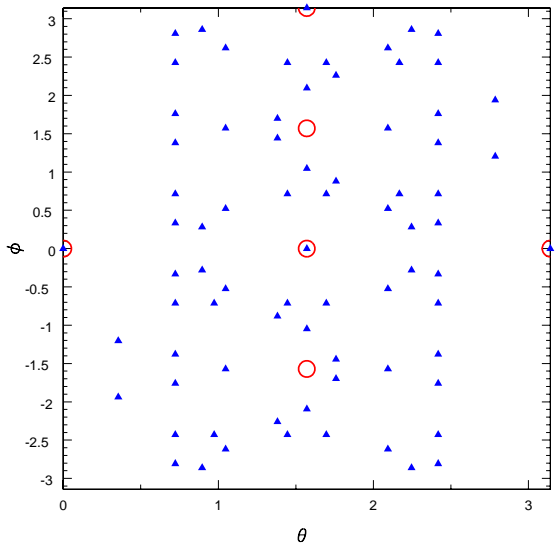}
	\caption{$\theta$ and $\phi$ are the standard angles in
 spherical coordinates, giving the orientations of a chosen reference
 vector in each tile of a level 7 Q-set (blue triangles), 
and the analagous quantities for a grid (red open circles). }
	\label{fig:angles}
\end{figure}

\subsection{Isotropy}
\label{sec:isotropy}
One of the nice features of the Q-set is, as we 
have underlined, that it is isotropic. This result applies, however, 
in the limit of an infinite number of iterations of the algorithm we 
have described. It it interesting to quantify the degree of 
isotropy of a finite level tiling.  To do so we show in 
Figure \ref{fig:angles} a plot giving (blue triangles) the distinct 
directions defined by the tiles of the quaquaversal tiling used to 
construct the Q-set with $N=7$. The directions are those of the 
shortest axis of each tile with respect to the orientation of the mother tile.
Also shown (red open circles) is an analagous characterisation of a
grid, which gives just the six orientations of the vectors 
pointing to nearest neighbour sites.  

\begin{figure}[htb]
	\centering
	\includegraphics[width=0.48\textwidth]{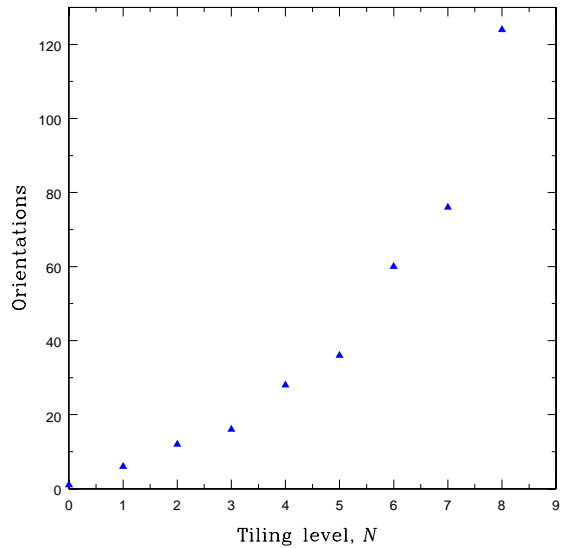}
	\caption{Number of distinct orientations as a function 
of level of tiling.}
	\label{fig:orientations}
\end{figure}

Figure \ref{fig:orientations} shows, on the other hand, the dependence
of the number of  distinct orientations, in the $(\theta,\phi)$-plane, 
on the level of the tiling. The degree of isotropy grows very rapidly
with increasing $N$.

\section{CONCLUSIONS}

We have studied an alternative to the standard pre-initial conditions
for N-body simulations of cosmological structures.  The standard {\em
grid} has strong orientations, and the {\em glass} is poorly defined
and computationally expensive to create. We have therefore considered
a particle distribution created starting from an equal volume
tiling of space, called the {\em quaquaversal tiling}. The 
particle distribution is
trivial to create, has virtually no orientation (is statistically
isotropic), and has rapidly vanishing large scale power-spectrum. We
provide a C-code for the generation of these structures on {\tt
http://krone.physik.unizh.ch/{}\~{}hansen/qua/}.

\acknowledgments SHH is supported by the Swiss National Foundation. MJ
is indebted to J. Lebowitz for the essential references on tilings,
and to C. Radin for subsequent useful exchanges. OA would like to
thank A.\,B. Romeo for valuable discussions. We also thank
T. Baertschiger, A.  Gabrielli, B. Jancovici, A. Knebe, A. Macci\`o,
B. Marcos, J. Peacock, F. Sylos Labini and S. Torquato for many useful
discussions on related issues.  We thank the anonymous referee for
very useful remarks and suggestions.

\end{document}